\documentstyle[12pt]{article}
\catcode`\@=11
\@addtoreset{equation}{section}

\def\blankcite{\@ifnextchar [{\@tempswatrue\@blankcitex}{\@tempswafalse\@blankcitex[]}}
\def\@blankcitex[#1]#2{%
  \let\@citea\@empty
  \@blankcite{\@for\@citeb:=#2\do
    {\@citea\def\@citea{,\penalty\@m\ }%
     \edef\@citeb{\expandafter\@iden\@citeb}%
     \if@filesw\immediate\write\@auxout{\string\citation{\@citeb}}\fi
     \@ifundefined{b@\@citeb}{{\reset@font\bfseries ?}%
       \G@refundefinedtrue\@latex@warning
       {Citation `\@citeb' on page \thepage \space undefined}}%
     {\hbox{\csname b@\@citeb\endcsname}}}}{#1}}
\def\@blankcite#1#2{{#1\if@tempswa , #2\fi}}

\catcode`@=12

\newfont{\mybb}{msbm10 scaled 1200}

\newcommand{\ind}{\hspace{.5cm}}

\newcommand{\be}{\begin{equation}}
\newcommand{\ee}{\end{equation}}
\newcommand{\bea}{\begin{eqnarray}}
\newcommand{\eea}{\end{eqnarray}} 
\newcommand{\nn}{\nonumber}

\newcommand{\m}{{\scriptscriptstyle -}}
\newcommand{\p}{{\scriptscriptstyle +}}
\newcommand{\sfrac}[2]{{\textstyle \frac{#1}{#2}}}
\newcommand{\intl}{\int\limits_{-L}^L}

\newcommand{\bra}{\langle}
\newcommand{\ket}{\rangle}

\newcommand{\real}{\mbox{\mybb R}}

\newcommand{\integer}{\mbox{\mybb Z}}

\newcommand{\pad}[2]{\frac{\partial #1}{\partial #2}}
\renewcommand{\pb}[2]{ \{#1, #2\} }
\newcommand{\vc}[1]{\mbox{\bf #1}}

\newcommand{\tr}{\mbox{tr}}
\newcommand{\FP}{\mbox{FP}}
\newcommand{\ad}{\; \mbox{ad}}

\newcommand{\CA}{{\cal A}}
\newcommand{\CG}{{\cal G}}
\newcommand{\CO}{{\cal O}}
\newcommand{\CM}{{\cal M}}
\newcommand{\CC}{{\cal C}}
\newcommand{\CQ}{{\cal Q}}
\newcommand{\CH}{{\cal H}}
\newcommand{\CL}{{\cal L}}

\renewcommand{\thefootnote}{\alph{footnote}}

\begin{document}

\thispagestyle{empty}

\hfill \parbox{4cm}{TPR-95-30 \\
                    hep-th/9604018}
\vspace{1cm}

\begin{center}
{\huge   Hamiltonian    Formulations    of   Yang-Mills   Quantum
Theory  and the Gribov Problem}{\Large  \footnote{Notes  based on
talks  given at the {\sl 5th Meeting  on Light-Cone  Quantization
and Non-Perturbative  QCD}, Regensburg, Germany, June 1995 and at
{\sl 35.~Internationale Universit\"atswochen  f\"ur Teilchen- und
Kernphysik}, Schladming, Austria, March 1996}}

\vspace{1.5cm}

{\large 
T.~Heinzl\footnote{e-mail: thomas.heinzl@physik.uni-regensburg.de} \\     
Universit\"at Regensburg \\   
Institut   f\"ur  Theoretische   Physik,  \\  
93053 Regensburg, FRG }

\end{center}

\vspace{1.5cm}

\begin{abstract}
We review the status of quantising  (non-abelian)  gauge theories
using      different      versions      of      a     Hamiltonian
formulation  corresponding  to Dirac's instant  and front form of
dynamics, respectively.  In order to control infrared divergences
we work in a finite spatial volume, chosing a torus geometry  for
convenience.   We  focus  on the  determination  of the  physical
configuration  space  of  gauge  invariant  variables  via  gauge
fixing.   This  naturally  leads  us to the  issue  of the Gribov
problem. We discuss it for different gauge choices, in particular
finite volume modifications of the axial gauge.  Conventional and
light-front quantisation are compared and the differences pointed
out. 
\end{abstract}

\vfill
\newpage

\renewcommand{\thefootnote}{\arabic{footnote}}
\setcounter{footnote}{0}
\setcounter{page}{2}

\section{Introduction}

\vspace{1cm}

The  quantisation  of non-abelian  gauge  fields  \cite{YM54}  is
notoriously difficult. The reason for this is the gauge symmetry:
at the level of, say, the Lagrangian a gauge theory is formulated
in terms of redundant degrees of freedom which change under gauge
rotations.  The  Lagrangian  itself  is invariant  such that  the
original  and the rotated configurations  correspond  to the same
physics.   Thus they should  be identified.   In a slightly  more
formal way this can be formulated as follows: Denote by $\CA$ the
total  configuration  space  of the gauge theory,  {\it i.e.}~the
space  of all gauge  fields  $A \in  \CA$.   The space  $\CA$  is
unphysical,  because it is ``too big'':  it contains  an infinite
number   of  gauge  equivalent,   hence   physically   identical,
configurations.   If $U(x)  \in \CG$  is an element  of the gauge
group  $\CG$,  and the gauge potential  $A \in \CA$, then a gauge
equivalent field is obtained by the gauge transformation

\be
A \to  U^{-1}  A  U - \frac{i}{g}  U^{-1}\partial  U
\equiv {^U\!\!}A \; , 
\ee

where $g$ denotes the coupling constant.   In order to get rid of
the gauge copies of the field $A$, one identifies  all of them to
form a single object (equivalence class) called the orbit $\CO_A$
of $A$,

\be
\CO_A \equiv \{{^U\!\!}A : U \in \CG \} \; .
\ee

A gauge transformation  thus leads from one point  on an orbit to
another  point on the {\em same} orbit,  so, by construction,  it
can never leave an orbit.  Two gauge fields on the same orbit are
called  gauge equivalent.   The set of all orbits is the physical
configuration  space  $\CM$.   In mathematical  terms this can be
written as

\be
\CM = \CA / \CG \; , 
\label{PHYSCONF}
\ee

thus $\CM$ is the space  where all gauge equivalent  fields  have
been   ``divided   out''.    Loosely   speaking   one   has   the
identity\footnote{Formula   (\ref{TIMES})  does  not  hold  in  a
mathematically  correct  sense  because  it says that $\CA$  is a
trivial $\CG$-bundle over $\CM$.  That this is not generally true
is the mathematical content of the Gribov problem (see below).}

\be
\CA = \CM \times \CG \; ,
\label{TIMES}
\ee

which states that the total configuration space $\CA$ consists of
the gauge invariant, physical variables forming $\CM = \CA / \CG$
and the gauge variant or ``cyclic''  variables  representing  the
gauge group $\CG$.   The main problem  with gauge theories  is to
disentangle  these  two  types  of variables  and make  the gauge
symmetry explicit by eliminating  the redundant  variables.   The
resulting  theory would then be formulated  in terms of the gauge
invariant  degrees  of freedom  only.   This  is very  much  like
formulating  a rotationally  invariant  system  in terms  of  the
radial variable only, after eliminating  the cyclic dependence on
the angles.  In a gauge theory, the problem is to properly divide
the variables into ``radial'' and ``angular'' ones.

\ind
Technically,  the existence of redundant degrees of freedom shows
up    through    the    appearance    of    {\em    constraints},
{\it i.e.}~non-dynamical   equations   between   the  field   variables
\cite{Ber49}.   In the Dirac formulation for constrained  systems
\cite{Dir50,  Dir64},  one finds at the first stage the vanishing
of the momentum  conjugate  to the  field  component  $A_0$  (the
``primary constraint'').   However, the real, physical constraint
is not this but a secondary one, Gauss's law, which is intimately
connected with gauge invariance as will be clarified later. It is
this constraint that has to be solved in one or the other way. In
other words, it has to be guaranteed  that Gauss's law holds once
and forever. Then gauge invariance is essentially preserved.

\ind
The formal expression (\ref{PHYSCONF}) is of no practical use for
finding  the  physical  configuration  space  $\CM$.   There  are
basically  two methods to deal with the abundance of variables in
a gauge theory and ``divide out'' the gauge group.  In the first,
one tries to construct  explicitly  gauge invariant variables out
of the original gauge fields and formulate the whole gauge theory
in terms of only these. This was pioneered by Mandelstam both for
QED  \cite{Man62}  and QCD \cite{Man68}.   In this  approach  one
trades   the  gauge   potentials   for  Wilson   loop   variables
\cite{Wil74} which are traced, path ordered exponentials,

\be
W_\gamma [A]  \equiv  \frac{1}{N}  \mbox{tr}  P \exp \oint_\gamma
dx^\mu A_\mu (x) \; ,
\ee

with $\gamma$ a closed path in space-time,  $P$ the path-ordering
operator   along   $\gamma$   and   $N$  the  dimension   of  the
representation  of the gauge  group.   The gauge  theory  is thus
formulated   in  ``loop  space'',  which,  unfortunately,   still
contains  redundant  degrees of freedom,  as can be seen from the
appearance of the ``Mandelstam constraints'' \cite{Man68, Man79}.
Only recently there has been some progress in solving these under
special circumstances \cite{Lol93}.

\ind
The  second  approach,  which  will  be  persued  here,  aims  at
identifying  $\CM$  via gauge  fixing,  {\it i.e.}~by  finding  a
subset  of $\CA$ that is isomorphic  to $\CM$  as explained  {\it
e.g.}~in  \cite{FSS94}.   Such a subset  is called  a fundamental
modular  region  (FMR)  \cite{vBa92}.   In order  to find it, one
imposes  conditions  on the gauge  potentials  $A$ such that  one
selects  a representative  $\bar A$ on each orbit.  Ideally,  one
would like to find a gauge condition $\CC [A] = 0$, such that the
physical  configuration  space  $\CM$ can be identified  with the
subset

\be
\Gamma \equiv \{A \in \CA : \CC [A] = 0 \} \; .
\label{GAUGEFIX}
\ee

Two  requirements  have  to be fulfilled  by an admissible  gauge
condition :

(I) {\em Existence:}

It  should   select   a  representative   on  {\em  any}   orbit;
this means that for any $A \in \CA$ there should be a solution $U
\in \CG$ of the equation

\be 
\bar A = {^U\!\!}A \; ,
\ee

with the representative $\bar A$ obeying the gauge condition $\CC
[\bar  A] = 0$.  Then any gauge field  $A$ can be transformed  to
this gauge.

(II) {\em Uniqueness:}

There should be only {\em one} representative  obeying  the gauge
condition  on each orbit.   If, on the other hand,  there are (at
least)  two gauge  equivalent  fields,  $\bar  A_1$, $\bar  A_2$,
satisfying  the  gauge  condition,  the gauge  is said  to be not
completely fixed.  There is a ``residual gauge freedom'' given by
the gauge transformation $V$ connecting the copies,

\be 
\bar A_2 = {^V\!\!}\bar A_1 \; .
\ee

It turns out that in the non-abelian case conditions (I) and (II)
cannot  be satisfied  using a simple  equation  $\CC [A] = 0$ for
gauge fixing \cite{Gri78, Sin78}. Thus, in order to eliminate the
copies  and  uniquely  specify  the  gauge,  one  has  to  choose
appropriate   conditions   in  addition  to  the  original   one.
Generically,  the former cannot be written  in terms of equations
but are of non-holonomic  nature  (like {\it e.g.}~inequalities).
It is obvious  that the fields  satisfying  (I) and (II)  form  a
subset $\Lambda  \subset \Gamma$ which constitutes  the FMD to be
identified  with the physical  configuration  space $\CM$.  So we
have the relations

\be 
\CM \cong \Lambda \subset \Gamma \subset \CA \; .
\ee

\ind
In the abelian case, {\it i.e.}~for QED, the gauge fixing program
is  straightforward   and  one  can  find  $\CM$   in  terms   of
(\ref{GAUGEFIX}).   The Coulomb gauge, $\nabla \cdot \vc{A} = 0$,
for example, eliminates the longitudinal  photons and one is left
with the physical, transverse ones. These correspond to the gauge
invariant, ``radial'' variables. As stated above, for non-abelian
gauge fields, the situation is much more involved.   It was first
noted by Gribov in 1978, that the Coulomb condition  does not fix
the gauge uniquely  \cite{Gri78},  a fact that has been named the
``Gribov   problem''   thereafter.    Gribov's  result  has  been
generalised  to a wide number  of gauges by Singer  \cite{Sin78}.
Although  there has been a lot of work done on this subject since
then, (see \cite{vBa92}  and references  therein),  it is fair to
say that the problem has still not been quite resolved.   This is
particularly true for gauge choices other than the Coulomb gauge.

\ind
So far we have not distinguished  between  the classical  and the
quantum   theory  as  the  remarks  above  apply  to  both.   The
quantisation  of  non-abelian  gauge  theories  is  a problem  in
itself, the features  discussed  above, however,  play a dominant
role.

\ind
There  are several  methods  of designing  a quantum  theory  for
non-abelian  gauge fields.  Yang and Mills in their original work
\cite{YM54}   used  a  Hamiltonian  formulation  derived  from  a
covariantly  gauge fixed Lagrangian \'{a} la QED, thus quantising
all four components of the gauge potential $A_\mu$.  The authors,
however,  did not address  the question  of finding  the physical
states, which in the covariant  (or Lorentz) gauge is non-trivial
as known from the Gupta-Bleuler  formulation  of QED \cite{GB50}.
This has been criticised  later by Schwinger who suggested to use
the  Coulomb  gauge  instead  \cite{Sch62a}.    By  insisting  on
relativistic  invariance, he obtained a Hamiltonian with a rather
non-standard  form for the kinetic energy.   The whole expression
looked  rather awkward  for practical  purposes.   Thus, in 1976,
Feynman stated that, ``Schwinger,  after a lot of hard work found
the Hamiltonian...,  but...progress  in this  direction  ceased''
\cite{Fey76}.    In  the  meantime,  people  had  abandoned   the
Hamiltonian  formulation  in favour of an alternative method, the
path integral.  The breakthrough  had been achieved with the work
of Faddeev  and Popov  \cite{FP67},  who  inserted  their  famous
determinant   in  the  path  integral   measure  leading  to  the
appearance  of ghost particles.   These served  to guarantee  the
unitarity  of Feynman  amplitudes,  explaining  earlier  work  of
Feynman \cite{Fey63} and De Witt \cite{DWi67}.   After that, path
integrals  were considered  to be {\em the} method  of quantising
non-abelian  gauge  fields,  and  the  shortest  route  to set up
Feynman rules and perturbation theory.

\ind
With the work of Gribov  \cite{Gri78},  however,  it became clear
that  the  Faddeev-Popov   method   must  be  incomplete   beyond
perturbation  theory.  This lead to a partial renaissance  of the
Hamiltonian  formulation,  which clearly  is not very well suited
for perturbation theory, but may offer a better understanding  of
(some) non-perturbative  phenomena.  This article comments on the
present state of affairs and discusses some new directions.

\ind
It is organised  as follows:   In Section  2, we formulate  gauge
theories (including  the Gribov problem) in the simple setting of
quantum mechanics.   Non-abelian  gauge fields are introduced  in
Section 3, where we shortly touch upon the Coulomb gauge and then
proceed with axial gauges in more detail.  In this section we use
the conventional equal-time quantisation based on Dirac's instant
form of relativistic  dynamics.   The last section  is devoted to
what is called light-cone quantisation  (originating  from Dirac's
front form) applied to Yang-Mills theory.

\vspace{1cm}

\section{Quantum Mechanical Gauge Systems}

\vspace{1cm}

The gauge theories usually considered are (local) field theories,
which  genuinely   have  an  infinite   number   of  degrees   of
freedom\footnote{An  exception  from  this  rule  are topological
field theories \cite{BBR91},  among them Yang-Mills  theory in $d
= 1+1$ defined on a compact manifold \cite{Mig75}.}.   The aspect
of gauge symmetry,  however, is not a field theoretic feature and
can be equally well discussed for systems with only a few degrees
of freedom.  Many of the general problems can be discussed  there
in a simpler  setting  \cite{CL80,  Pro82,  CT87a}. 

\ind
Consider  for example a free (point)  particle  moving on a plane
\cite{CL80, CT87a}. Its configuration space is

\be
\CQ =  \{ (q_1, q_2) \in \real^2 \} \; ,
\ee

and its Lagrangian

\be
L(p,q)  =  p_1  \dot  q_1  + p_2  \dot  q_2  - \frac{p_1^2}{2}  -
\frac{p_2^2}{2} \; .
\label{QMLAG}
\ee

From  the  kinetic  term  we can read  off the canonical  Poisson
brackets,

\be
\pb{q_i}{p_j} = \delta_{ij} \; , \quad i,j = 1,2 \; 
\ee

As an additional condition we impose that the angular momentum of
the particle should be vanishing,

\be
G \equiv q_1 p_2 - q_2 p_1 = 0 \; ,
\label{QMGAUSS}
\ee

a constraint  which we call ``Gauss's law'' for reasons to become
clear in a moment.  The angular momentum $G$ is the generator  of
rotations  around the origin of the plane, and we interpret it as
the generator of time-independent  gauge transformations  so that
the gauge group is $O(2)$.  Any motion of the particle  violating
condition (\ref{QMGAUSS}) is thus gauge-variant, {\it i.e.}~unphysical.
Equations (\ref{QMLAG}) and (\ref{QMGAUSS}) completely define our
mechanical gauge system.

\ind
On the classical  level, the gauge invariance  of the Hamiltonian
$H=  \vc{p}^2  /2$ is expressed  by the vanishing  of the Poisson
bracket

\be
\pb{G}{H} = 0 \; ,
\ee

which simultaneously states the invariance of the Hamiltonian and
the time  independence  of the angular  momentum.   In this  very
simple example it is clear that the orbits are circles around the
origin of radius

\be
r = \sqrt{q_1^2 + q_2^2} \; ,
\label{RADIUS}
\ee

the points  of which  are transformed  into each other  via gauge
rotations  by some  angle  $\phi$.   Note that  the origin  is an
exceptional  point.   There, the orbits degenerate  into a single
point  which  is not transformed  under  the action  of the gauge
group.   The origin  is thus a fixed point and the action  of the
group  is not free.   The orbit corresponding  to the origin  has
dimension zero whereas all others have dimension one.

\ind
In this example, the gauge invariant variable is obviously  given
by the  radius  $r$,  as the  radial  motion  is associated  with
vanishing angular momentum, whereas the angle

\be
\phi = \tan^{-1} (q_2 / q_1)
\ee

is gauge-variant or ``cyclic''. This can be explicitly checked by
calculating the Poisson brackets

\be
\pb{r}{G} = 0 \; , \quad \pb{\phi}{G} = 1 \; .
\ee

Thus,  the  angle  $\phi$  and  the  angular  momentum   $G$  are
canonically  conjugate.   As $r$ is the gauge invariant variable,
the physical configuration space is

\be
\CM = \real^2 / O(2) = \real_0^+ = \{r \in \real :  r \ge 0 \} \;
. 
\ee

The quantisation is straightforward.  The Hilbert space of states
consists  of wave  functions  $\Psi  (r)$  being  independent  of
$\phi$ which expresses  their gauge invariance.   The Hamiltonian
$H$  acting   upon  these   is  just  the  radial   part  of  the
two-dimensional Laplacian,

\be
H = -\frac{1}{2} \, \frac{1}{r} \pad{}{r} r \pad{}{r} \; .
\label{QMHAM}
\ee

Let us now try to obtain the same results  via gauge fixing.  The
reader  should imagine  a situation  where it is too hard to find
the gauge invariant  variables, or the ones found are too awkward
to use.  In this case, gauge fixing offers a pedestrian's  way to
the physical configuration space.

\ind
For the case at hand, we choose the gauge 

\be
\chi (q_1 , q_2) \equiv q_2 = 0 \; .
\label{QMGAUGE}
\ee

This gauge exists,  since for any point with coordinates  $(q_1 ,
q_2)$ one finds a gauge rotation $U[\phi]$  transforming  it to a
point $(\bar q_1 , 0)$ with vanishing  $q_2$,  {\it i.e.}~on  the
horizontal  axis.   The angle  of rotation  is given  by $\phi  =
\tan^{-1} (q_2 / q_1)$.  However, the gauge fixing is not unique,
since for any point $(\bar q_1 , 0)$ on the $q_1$-axis there is a
copy  $(-\bar  q_1  , 0)$  obtained  via  rotation  by $\pi$  or,
equivalently, a reflection at the origin. Accordingly, there is a
residual  gauge  invariance,   not  fixed  by  the  gauge  choice
(\ref{QMGAUGE}), given by the reflections $q_1 \to - q_1$.

\ind
This  observation  can  be  cast  in a more  mathematical  frame.
Consider the Faddeev-Popov (FP) ``matrix'' \cite{FP67}

\be
\FP \equiv \pb{\chi}{G} = q_1 = \FP (q_1) \; ,
\ee

which in this simple example degenerates to a single real number.
Thus, its modulus is the FP determinant,

\be
J \equiv |\det \FP | = |q_1 | \; .
\ee

Both quantities,  however, are not constant, but dependent on the
coordinate $q_1$.  The point where the determinant  is vanishing,
$q_1 = 0$, is seen to be a Gribov horizon separating the two gauge
equivalent regions $q_1 > 0$ and $q_1 < 0$.  For a complete gauge
fixing  one  has  to demand  in addition  that  $q_1$  should  be
nonnegative,  $q_1 \ge 0$.  This non-holonomic  condition defines
the FMD of the problem.  Again we have consistently obtained $\CM
= \real^+$,  and we can identify $q_1$ with the radius $r$, which
follows  from the definition  (\ref{RADIUS})  upon inserting  the
gauge   condition   (\ref{QMGAUGE}).    This  leads   to  another
interpretation  of the FP determinant:  it is the Jacobian of the
transformation  from the original, cartesian coordinates  $(q_1 ,
q_2)$   to  the  curvilinear   polar   coordinates   $(r,  \phi)$
\cite{CL80}, where, {\em after complete gauge fixing}, the radial
variable $r$ can be identified  with the cartesian variable $q_1$
restricted to the FMD.

\ind
Let us discuss  the gauge fixing procedure  in the quantum theory
in more detail. Before gauge fixing, the quantum Hamiltonian is

\be
H = \frac{1}{2} (p_1^2 + p_2^2) \; ,
\ee

with the canonical commutators 

\be
[ q_i , p_j ] = i \delta_{ij} \; .
\ee

In order not to get a contradiction  to the commutation relations
we  can  only  impose  Gauss's  law  weakly,  {\it  i.e.}~on  the
(physical) states,

\be
G |\Psi \ket = 0 \; .
\ee

The Schr\"odinger wave functions are defined as usual,

\be
\Psi (q_1 , q_2) = \bra q_1 , q_2 | \Psi \ket \; ,
\ee

The momentum operators acting upon them are

\be
p_i = -i \pad{}{q_i} \; .
\ee

This  is  the  usual  Schr\"odinger  quantisation  for  cartesian
coordinates.   Gauge  fixing  in the quantum  theory  is done  by
demanding  that the physical states (satisfying  Gauss's law) are
given by a projection onto $q_2 = 0$,

\be 
\Psi_\chi   (q_1)   \equiv  \left.    \bra  q_1  ,  q_2  |  \Psi  \ket
\right\vert_{q_2 = 0} = \bra q_1 , 0 | \Psi \ket \; .
\ee

On the gauge fixed  physical  states  the momentum  $p_1$ acts as
the usual differential  operator,  $p_1 = -i \partial  / \partial
q_1$, but, due to Gauss's law,

\be
p_2 | \Psi \ket = q_1^{-1} q_2 p_1 |\Psi \ket \neq -i \pad{}{q_2}
|\Psi \ket \; . 
\ee

This yields \cite{CT87a}

\bea
\bra q_1 , 0 | \, p_2^2  \, | \Psi \ket &=& \bra q_1 , 0 | \, p_2
(q_1^{-1}  q_2  p_1)  |\Psi  \ket  \nn  \\ 
&=& \bra q_1 , 0 | -i q_1^{-1} p_1 + (q_1^{-1}  q_2 p_1)^2 | \Psi
\ket \nn \\
&=& - \frac{1}{q_1} \pad{}{q_1} \Psi_\chi (q_1) \; , 
\eea

and thus the Hamiltonian 

\be
H = - \frac{1}{2} J^{-1} (q_1) \pad{}{q_1} J(q_1) \pad{}{q_1} \;
\label{QMHAMFIX}
\ee

acting on gauge invariant states.  As expected, it coincides with
(\ref{QMHAM})  upon identifying  $J(q_1)  = |q_1|$ with $r$.  The
Hamiltonian  (\ref{QMHAMFIX})  is hermitean  with respect  to the
scalar product

\be
\bra  \Psi | \Phi \ket = \frac{1}{2\pi}  \int d^2 q \Psi^*  (q)
\Phi (q)  = \int_0^\infty  dq_1 q_1 \Psi_\chi^*  (q_1) \Phi_\chi
(q_1) \; ,
\ee

where  we have  divided  by the  ``volume  of the gauge  group'',
$2\pi$. The associated Hilbert space of physical states is thus

\be
\CH = {\mybb L}^2 (\real_0^+, J(q_1) dq_1 ) \; .
\ee

The global  properties  of the configuration  space  $\real_0^+$,
which is a half-space  with a boundary, affect the wave functions
in a peculiar  way.  One can imagine  the configuration  space as
being a semi-infinite  potential  well, so that the wave function
can not penetrate the barrier to negative values of $q_1$ or $r$.
This necessitates the boundary condition

\be 
\Psi_\chi (q_1 = 0) = 0 \; .
\ee

The wave  function  is thus  repelled  from the origin.   Another
way to see this is to simplify  the measure in the scalar product
by absorbing  a factor  $J^{1/2}$  into  the wavefunction.   This
results in a rescaled Hamiltonian 

\be
\bar    H    =    J^{-1/2}    H    J^{1/2}    =   -   \frac{1}{2}
\frac{\partial^2}{\partial q_1^2} - \frac{1}{8 q_1^2} \; ,
\ee

which  has  a  standard  form  for  the  kinetic  energy  but  an
additional ``effective potential''  term $1/ 8q_1^2$, reminiscent
of  the  centrifugal  barrier  in  the  hydrogen  atom.  The  new
Hamiltonian is hermitean on ${\mybb L}^2 (\real_0^+ , dq_1)$.

\ind 
Let us close this section with a mathematical  remark.  It should
be noted that one cannot write the configuration  space $\real^2$
as a trivial bundle over $\real_0^+$,  {\it i.e.} as a product of
the physical configuration space with the gauge group,

\be 
\real^2 \neq O(2) \times \real_0^+ \; ,
\ee

because  of the singularity  at the fixed point $r = 0$.  This is
tantamount  to saying  that the mapping  from cartesian  to polar
coordinates,

\be
(q_1 , q_2 ) \mapsto (r , \phi) \; ,
\ee

is not one-to-one.   At the origin,  $r=0$, the associated  angle
$\phi$ is not defined.  This phenomenon, the Gribov problem, does
equally occur within Yang-Mills theory, which is our next topic.

\vspace{1cm}

\section{Hamiltonian Yang-Mills Theory: Instant Form}

\vspace{1cm}

\subsection{General Remarks}

\vspace{1cm}

In the instant  form of relativistic  dynamics  \cite{Dir49}  one
choses  the ordinary  Cartesian  time $t = x^0$  as the parameter
describing   the  causal  evolution   of  a  given  system.   The
Hamiltonian   then   propagates   a  state   according   to   the
Schr\"odinger equation,

\be
H \vert \psi \ket =  i \pad{}{t} |\psi \ket \; .
\ee

This  choice  of time  is most  common  and  coincides  with  the
non-relativistic  one, which is unique due to Galileian causality.
Einstein causality, however, leaves room for alternative choices,
one of which,  leading  to the front-form  of dynamics,  will  be
discussed in the next section.

\ind
First of all, let us explain our basic conventions  and notations
for gauge  field  theory.   Throughout  this paper  we will use a
finite spatial  volume as an infrared  regulator,  following  the
approach  pioneered  by  `t  Hooft  \cite{tHo79}   and  L\"uscher
\cite{Lue83}.  We let $-L \le x_i \le L$, $i = 1,2,3$, and impose
periodic boundary conditions (pBC) on the gauge fields,

\be
A_i (x_j = -L) = A_i (x_j = L) \; , \quad i,j = 1,2,3 \; .
\ee

This results in a torus geometry for three-space. We write the YM
Lagrangian in the following form \cite{FS80, FJ88}

\bea
\CL &=& \CL_{kin} - H + A_0^a G^a \nonumber \\
&\equiv&  E_i^a  \dot A_i^a - \frac{1}{2}  (E_i^a  E_i^a  + B_i^a
B_i^a) + A_0^a D_i^{ab}E_i^b \; ,
\label{LYM}
\eea

where $\dot A_i^a$ denotes the time derivative  of the field (the
velocity), $H$ the Hamiltonian and $G^a$ the Gauss operator.  For
simplicity,  we choose $SU(2)$ as the gauge group and will switch
freely  between component  and matrix notation  for the gauge
fields,  $A_i = A_i^a  T^a$, the chromo-electric  fields,  

\be
E_i = E_i^a T^a = \dot A_i - D_i A_0  \; ,
\ee

and the chromo-magnetic fields

\be
B_i = B_i^a T^a = \epsilon_{ijk} (\partial_j + ig A_j ) A_k \; .
\ee

As a basis  for the Lie algebra  $su(2)$  we have  chosen  $T^a =
\tau^a/2$, with $\tau^a$ the usual Pauli matrices.  The covariant
derivative is

\be
D_i^{ab} = \partial_i \delta^{ab} + g \epsilon^{abc} A_i^c \; ,
\ee

or in $su(2)$ matrix notation,

\be
D_i = \partial_i  + ig [A_i , \quad ] \equiv \partial_i  + ig \ad
A_i \; \quad . 
\label{CODERIV}
\ee

From the kinetic term of (\ref{LYM}) we read off \cite{FJ88}  the
canonical commutators,

\be
[A_i^a  (\vc{x}  ,  t)  ,  E_j^b  (\vc{y}  ,  t)  ] = \delta_{ij}
\delta^{ab} \delta (\vc{x} - \vc{y} ) \; .
\ee

Thus,  gauge fields  $A_i$ and chromo-electric  fields  $E_i$ are
canonically  conjugate  and generalise  the cartesian coordinates
$q$, $p$ of Section 2.  It is therefore straightforward to set up
a  functional   Schr\"odinger   picture   with  wave  functionals
$\Psi[A_i]$  replacing  the wave functions  of quantum mechanics.
The canonical momenta act as

\be
E_i^a \Psi [A_i] = - i \frac{\delta}{\delta  A_i^a} \Psi [A_i] \;
,
\ee

from which one can infer the commutation  properties of the Gauss
operator

\be
[G^a (\vc{x})  , G^b (\vc{y})  ] = i \epsilon^{abc}  G^c (\vc{x})
\delta ( \vc{x} - \vc{y} ) \; , 
\label{GAUSSCOMM} 
\ee

which yields  a representation  of the Lie algebra  $su(2)$.   As
emphasized  by Jackiw \cite{Jac80}, the field theory discussed so
far might be totally consistent  but differs from the {\em gauge}
field theory defined by additionally  imposing Gauss's law on the
(physical) states,

\be
G^a \Psi [A_i] = 0 \; .
\label{INFGAUSS}
\ee

The  Gauss  operator  generates  time-independent,  infinitesimal
gauge transformations according to

\be
[A_i^a  ,  G^b]  =  i D_i^{ab}  = i \delta^{ab}  \partial_i  + ig
\epsilon^{abc}A_i^c \; , 
\label{TRAFO}
\ee

where we have omitted the spatial arguments for convenience.  The
first term on the {\it r.h.s.} of (\ref{TRAFO})  corresponds to a
translation  in ordinary space, the second, non-abelian one, to a
rotation in color space.  Chromo-electric  and -magnetic  fields,
however, transform homogeneously,

\bea
[E_i^a , G^b] &=& ig \epsilon^{abc} E_i^c \; , \\
\lbrack B_i^a , G^b] &=& ig \epsilon^{abc} B_i^c \; .
\eea 

The Hamiltonian is (gauge-)invariant under  (\ref{TRAFO}),

\be
[H, G^a ] = 0 \; ,
\ee

thus  Gauss's  law is consistent  with  the equations  of motion.
Finite gauge transformations are generated by the operator

\be
\Omega  [\theta]  = \exp  -i \int  d^3x E_i^a  (\vc{x})  D_i^{ab}
\theta^b (\vc{x}) \; , \ee

such that the transformed field is given by

\be
\Omega  A_i  \Omega^{-1}  = U^{-1}A_i  U - \frac{i}{g}
U^{-1} \partial_i U \equiv {^U\!\!}A_i \; , 
\ee

where $U = U[\theta]$ is defined as

\be
U (\vc{x}) = \exp i g \theta^a (\vc {x}) T^a \in SU(2) \; .
\ee

The finite version of Gauss's law, (\ref{INFGAUSS}), is then

\be
\Psi [{^U\!\!}A_i] = \Psi [A_i] \; ,
\ee

expressing  the gauge invariance  of the physical states.  Let us
now  try  to find  the physical  configuration  space  via  gauge
fixing.  Before we discuss any particular  example,  some general
remarks are in order. We denote the gauge fixing condition by

\be
\chi[A_i] = 0 \; 
\ee

and imagine a coordinate transformation

\bea
A_i    &\to& (A_i^* , \chi) \; , \\
\Pi_i  &\to& (\Pi_i^* , \pi) \; , 
\eea

where $\Pi_i = E_i$ is canonically  conjugate to $A_i$, and $\pi$
to  $\chi$.  The  gauge  invariant  canonical  pair  is  $(A_i^*,
\Pi_i^*)$.   The ideal case seems to be that $\pi$ coincides with
the Gauss operator  $G$, and that the gauge condition  $\chi$  is
conjugate  to it.  Gauss's  law would  then state  that  the wave
functionals  are independent of $\chi$ (being the analogue of the
angle $\phi$ of Section 2),

\be
G^a \Psi [A_i^*  , \chi] = -i \frac{\delta}{\delta  \chi^a}  \Psi
[A_i^* , \chi] = 0 \; ,
\label{IDEALGAUSS}
\ee

and we would have found the physical  configuration  space  $\CM$
consisting  of the $A_i^*$.   In this case  the FP matrix  is (in
symbolic notation) $\FP = \pb{\chi}{G}  = 1$, its determinant has
no zeros and the gauge fixing were complete.  Following  Gribov's
original observations \cite{Gri78}, Singer has proven that such a
scenario  cannot possibly be achieved \cite{Sin78}  within gauges
allowing for one-point compactification, $\real^4 \cup \{\infty\}
\cong  S^4$ \cite{Wei83},  so that the situation  is considerably
more  complicated.   This  can be easily  seen \cite{CT87a}  from
(\ref{IDEALGAUSS}):   if  the  Gauss  operator  could  really  be
represented   as  the  (functional)   differential   operator  of
(\ref{IDEALGAUSS}),    it   would   commute   with   itself,   in
contradiction    to   (\ref{GAUSSCOMM}).     Clearly,   no   such
contradiction  is present in the abelian case.  Similar arguments
have been given recently in \cite{LM94}.

\ind
Let us consider the FP matrix for arbitrary  gauge fixing $\chi$.
In terms of the original variables $A_i$ one finds

\be 
\FP = \pb{\chi}{G}_{\chi  = 0} = \left.  \frac{\delta \chi}{\delta
A_i} D_i \right\vert_{\chi = 0} \; .
\label{FPM1}
\ee

So,  if  $\chi$  is  linear  in  the  $A_i$,  the  FP  matrix  is
essentially the covariant derivative.  In terms of the new, gauge
fixed variables $A_i^*$, $\pi$, the FP matrix is \cite{FS80}

\be
\FP = \frac{\delta G}{\delta \pi} \; .
\label{FPM2}
\ee

From this last expression  one infers  that if the FP determinant
is non-vanishing,

\be
J = | \det \FP | = |\det \frac{\delta G}{\delta \pi} | \neq 0 \; ,
\ee

Gauss's law can be solved (locally) for the gauge variant momenta
$\pi$ by inverting the FP matrix. If this can be done explicitly,
one  is  left  with  the  gauge  invariant  variables  $(A_i^*  ,
\Pi_i^*)$.   It will turn out, however, that also this program is
beset with obstructions. In general, the FP matrix is a (partial)
differential operator.  Ideally, one would like to know its whole
spectrum.   Anyhow, in order to invert it, one has to specify its
domain together  with boundary  conditions.   This is another, if
technical, reason to work in a finite, definite geometry like the
torus.  Still, the FP matrix might be too complicated an operator
to be inverted exactly.

\ind
It should again be noted that, although the FP matrix was defined
within the classical theory in (\ref{FPM1}) and (\ref{FPM2}),  it
has a strong  impact on the quantum  theory.   Using a functional
Schr\"odinger  picture, a FP matrix that is only dependent on the
$A_i$, (which is the generic  case), acts as a pure number on the
functionals $\Psi [A_i]$. The remarks above thus equally apply to
the quantum  theory, when Gauss's  law is solved in the course of
defining  the quantum Hamiltonian  acting on the physical  states
via

\be 
H_\chi \Psi_\chi [A_i^*] = \bra A_i \vert H \vert \Psi \ket_{\chi
= 0} \; .
\ee

Let us now discuss some prominent examples of gauge fixing.

\vspace{1cm}

\subsection{The Coulomb Gauge}

\vspace{1cm}

The Coulomb gauge condition, which is the most natural one in the
abelian  case \cite{LM93},  demands  transversality  of the gauge
potentials,

\be
\chi [A_i] = \partial_i A_i = 0 \; .
\ee

Within  the Hamiltonian  formulation,  it is the most widely used
and therefore  best  studied  gauge  fixing  condition  (see  for
example \cite{vBa92,  Sch62a, CL80, FS82, vBa94}).  The existence
has been implicitly  proved  by Semenov-Tyan-Shanskii  and Franke
\cite{FS82}.   The gauge invariant  variables  are (like  in QED)
chosen  to be the transverse  gauge potentials,  $A_i^* = A_i^T$,
such that the physical  states  are $\Psi  = \Psi [A_i^T]$.   The
canonical  momentum,  the electric  field,  is decomposed  into a
longitudinal and transverse part \cite{Sch62a},

\be
E_i = E_i^T + E_i^L \equiv E_i^T + \partial_i \phi \; ,
\ee

where a scalar potential $\phi$ has been introduced.   The latter
is canonically conjugate to the gauge condition, 

\be
\pb{\chi}{\phi} = 1 \; ,
\ee

therefore  corresponds  to the gauge variant  momentum  $\pi$ and
should  be eliminated  via  Gauss's  law, (to  be imposed  on the
physical states),

\be
G\vert_{\chi = 0} = \partial_i D_i [A_i^T ] \phi + [A_i^T , E_i^T
] \equiv \FP \phi - \rho_c = 0 \; .
\ee

This  expression  might  be  called  the  ``non-abelian   Poisson
equation'' with the Laplacian replaced by the FP operator, $\FP =
\partial_i  D_i$,  and  the color-charge  density  of the gluons,
$\rho_c$.  As stated above, it can be solved for $\phi$ if the FP
determinant  $J = |\det  \FP|$ is non-vanishing.   The latter  is
difficult to check, since the FP matrix is a complicated  partial
differential  operator.  Its inverse is not explicitly  known and
can only be approximated  as a power series in the coupling  $g$.
Formally this means

\be
\FP^{-1} = \Delta^{-1} + O(g) \; ,
\label{FPPERT}
\ee

so that the perturbative  expansion can be viewed as an expansion
around the abelian  limit.  As a result,  the Hamiltonian,  which
depends  on $\FP^{-1}$  and $J$, can only  be written  down  in a
formal or a perturbative sense \cite{Sch62a, CL80, CT87a, Khr69}.

\ind
As  mentioned   above,   Gribov   was  the  first  to  point  out
\cite{Gri78}   that   for   large   fields,   {\it   i.e.}~beyond
perturbation   theory,   the  $\FP$  matrix  develops   vanishing
eigenvalues  on what  are now called  ``Gribov  horizons''.   The
(convex)  interior  of the first horizon,  the Gribov region $B$,
however,  still  contains  Gribov  copies  and  has to be further
restricted to form a FMD. This is most transparently discussed in
terms of a (Morse) functional  on $\CA$.  Its critical points are
the fields  $A_i$ satisfying  the Coulomb  gauge  condition,  its
relative  minima  constitute  the Gribov region $B$, its absolute
minima the FMD $\subset B$ \cite{FSS94, vBa92, FS82}.

\ind
In view of the above, in particular  (\ref{FPPERT}),  one can say
that the Coulomb  gauge  is well suited  for perturbation  theory
\cite{LM94}.   This  has been used {\it e.g.}~in  the Hamiltonian
proofs of asymptotic freedom \cite{Khr69} or in L\"uscher's small
volume  calculations  \cite{Lue83}.  To  go  beyond  perturbation
theory is possible but technically involved \cite{vBa92,  KvB88}.
It is therefore desireable  to look for and find alternatives  to
the Coulomb gauge formulation.

\vspace{1cm}

\subsection{Axial-Type Gauges}

\vspace{1cm}

Axial  gauges  have  been  introduced  independently   by  Kummer
\cite{Kum61}  and Arnowitt  and Fickler \cite{AF62}  in the early
sixties (for recent reviews see \cite{Lei87,  BNS91}).   They are
defined by the condition

\be
n \cdot A = 0 \; ,
\ee

where $n$ is a constant four vector.  Depending on whether $n$ is
time-like, space-like or light-like,  one refers to the temporal,
the axial or the light-cone gauge, respectively.  Their algebraic
simplicity  was -- and still is -- the main motivation  for using
them,  despite  their  lack  of explicit  covariance.   As the FP
operator, $\FP = n \cdot \partial$,  is field-{\em  independent},
the gauge is (apparently) ghost free \cite{Fre76}.   Accordingly,
the constraint  equations  are rather easy to solve, one does not
have to transform  to curvilinear  coordinates  and there  are no
ordering problems,  in contrast to the Coulomb gauge.  But it has
already been noted by Schwinger  in 1963 \cite{Sch63}  that there
is  a  huge  residual   gauge  freedom:    gauge  transformations
independent  of  $n  \cdot  x$  are  compatible  with  the  gauge
condition.  These transformations  correspond to zero modes (ZMs)
of the FP operator,  thus, $\det \FP = \det (n \cdot \partial)  =
0$.   As  pointed  out  by  Schwinger,  this  leads  to  infrared
divergences  in the Hamiltonian.   The solution  to the  infrared
problems is again to work in a finite spatial volume, $-L \le x_i
\le L$, and impose pBC (torus geometry). There is however a price
for this to be paid: the (strict) axial gauge does not exist on a
torus.   For the abelian  case, this has been first noted  by Yao
\cite{Yao64}, a student of Schwinger's. For the non-abelian case,
it is implicit  in the work of Batakis and Lazarides  \cite{BL78}
on the vacuum of finite-temperature  Yang-Mills theory, where pBC
naturally  arise  in the (imaginary)  time direction  so that the
temporal  gauge cannot be attained (see also \cite{Wei81}).   For
the light-cone  gauge,  the same  was first  noted  by Franke  et
al.~\cite{FNP81},  for space-like  axial gauges it is implicit in
\cite{Wei83,  BLS84} and explicitly stated in \cite{Pal86, Yab89,
Pal90}.

\ind
Let  us for the sake  of definiteness  consider  the (space-like)
axial  gauge,  $A_3 = 0$.  It turns out then, that configurations
$a_3 (x_1, x_2)$, which are independent  of $x_3$ and thus ZMs of
$\FP    =    \partial_3$,    cannot    be    gauged    away    by
allowed gauge transformations\footnote{These  are periodic  up to
an element  of the center  of the gauge  group  and thus preserve
periodicity  of the gauge  fields  \cite{Lue83}.   The center  of
$SU(2)$, for example, is $\integer_2 = \{\pm 1 \}$.}. Another way
to see this is the following.   Consider  the Wilson loop winding
around the torus,

\be
W [A_3] = \frac{1}{N} \tr P_3 \exp ig \intl dx_3 A_3 \; ,
\ee 

where $P_3$ denotes path ordering in $x_3$ direction.  The Wilson
loop  is  a  gauge-invariant,  dynamical  quantity.  In  general,
therefore,  $W[A_3]  \neq  1$, whereas  $W[A_3  = 0] = 1$.  If we
decompose $A_3$ into a ZM $a_3$ and a non-ZM $A_{33}$,

\be 
A_3 = A_{33}  + a_3 \; , \quad a_3 = \frac{1}{2L}  \intl dx_3 A_3
\; ,
\ee

one  can  only  gauge  away  $A_{33}$  but  has  to  retain  a ZM
$a_3^\prime$,  the gauge transform of $a_3$.  This corresponds to
the modified gauge condition

\be
\partial_3 A_3 = 0 \; . 
\label{MODAX}
\ee

Thus,  the  Fourier  transform  $\tilde  A_3  (\vc{k})$  is  zero
everywhere, except for $k_3 = 0$ \cite{Yao64}. One should also be
careful to realize that in the non-abelian case,

\be
W [A_3]  = \frac{1}{N}  \tr  P_3  \exp  ig \intl  dx_3  A_3  \neq
\frac{1}{N} \tr \exp [2igLa_3] \; ,
\label{ZMWILSON}
\ee

due  to  the  path-ordering  operation.   Also,  with  the  gauge
condition  (\ref{MODAX}), one still has to fix the residual gauge
freedom of gauge transformations independent of $x_3$.  There are
many possible  choices \cite{Yao64,  Pal86, Yab89, LNT94, KLS94}.
We will follow Palumbo \cite{Pal86,  Pal90}, and call this choice
henceforth  Palumbo gauge.  To define it, we introduce a momentum
lattice via Fourier decomposition into plane waves,

\be
e(\vc{n}  \cdot \vc{x}  ) = \exp \Bigl[i \pi  (n_1 x_1 + n_2 x_2 +
n_3 x_3)/ L \Bigr] \; .
\ee

For an arbitrary phase space variable, $F$, we write 

\bea
F (x_1,  x_2, x_3) &=& \sum_{n_1  , n_2 , n_3} \tilde  F (n_1 ,
n_2 , n_3) e(n_1 x_1 + n_2 x_2 + n_3 x_3 ) \nonumber \\
&=& \sum_{n_1  , n_2 , n_3 \atop n_3 \neq 0} \tilde  F (n_1 ,
n_2 , n_3) e(n_1 x_1 + n_2 x_2 + n_3 x_3 ) + \nonumber \\
&+& \sum_{n_1 , n_2 \atop n_2 \neq 0} \tilde F (n_1 , n_2 , n_3
= 0) e(n_1 x_1 + n_2 x_2 ) + \nonumber \\
&+& \sum_{n_1  \neq 0} \tilde F (n_1 , n_2 = 0 , n_3 = 0) e(n_1
x_1) + \nonumber \\
&+& \tilde F (n_1 = 0 , n_2 = 0,  n_3 = 0) \nonumber \\
&\equiv&  F_{3}  (x_1 , x_2 , x_3) + F_{2} (x_1 , x_2) + F_{1}
(x_1) + F_{0} \; .
\label{FIELDDECOMP}
\eea

In this way we have split  the function  $F$ into four components
$F_{r}$ depending on less and and less space coordinates $x_i$,

\be
F (\vc{x}) = \sum_{r=0}^3 F_{r} \; .
\ee

In coordinate space, the components $F_r$ read explicitly

\bea
F_3 (x_1 , x_2 , x_3 ) &=& F(x) - \frac{1}{2L}  \intl  dx_3  F(x)
\; , \\
F_2  (x_1   ,  x_2  )  &=&  \frac{1}{2L}   \intl   dx_3  F(x)   -
\frac{1}{(2L)^2} \intl dx_3 dx_2 F(x) \; , \\
F_1  (x_1  )  &=&  \frac{1}{(2L)^2}   \intl   dx_3  dx_2  F(x)  -
\frac{1}{(2L)^3} \intl d^3 x F (x)  \; , \\
F_0 &=& \frac{1}{(2L)^3} \intl d^3 x F (x)  \; .
\eea

All phase space  variables  (gauge  fields  and momenta)  will be
decomposed analogously. In the following, all summations over $r$
will be written  explicitly  to avoid confusion.   The components
$A_{ir}$, $r>0$, do not contain a ZM with respect to $x_r$,

\be
\intl dx_r A_{ir} = 0 \; , \quad r > 0 \; ,
\ee

and satisfy

\be
\partial_r A_{is} = 0 \; , \quad r > s \; .
\ee

The Palumbo gauge is then defined by the following conditions

\bea
A_{33} (x_1 , x_2 , x_3) &=& 0  \; , \nn \\
A_{22} (x_1 , x_2) &=& 0 \; , \nn \\ 
A_{11} (x_1) &=& 0 \; , 
\eea

which in shorthand notation is simply

\be
A_{rr} = 0 \; , \quad r = 1,2,3 \; .
\label{PALUMBO}
\ee

A last definition we need is 

\be
a_r \equiv \sum_{s=0}^{r-1} A_{rs} \; ,
\ee

representing  the sum of all ZMs with respect  to $x_r$ contained
in $A_r$, thus

\be 
\partial_r a_r = 0 \; .
\ee

Our next task is to show that the Palumbo  gauge  (\ref{PALUMBO})
exists.   This proof has not been given  in the original  work of
Palumbo \cite{Pal86, Pal90}, who simply assumed the existence. As
a consistency  check  he  solved  Gauss's  law  for  the  momenta
conjugate  to the $A_{rr}$  without redundant  degrees of freedom
being  left.   One can, however,  give a direct  existence  proof
\cite{Hei} by constructing a gauge transformation  $U$ that takes
an arbitrary configuration $A$ to Palumbo gauge, with

\be  
{^U\!\!}A_{rr}  = (U^{-1} A_r U - \frac{i}{g} U^{-1} \partial_r U
)_r = 0 \; . 
\ee

To do so, one has to perform  three  steps for $r = 1,2,3, $ such
that  $U = U_1 U_2 U_3 $, $\partial_r  U_s = 0$ for  $r>s$.   The
following table schematically depicts the procedure.

\begin{center}

\begin{tabular}{|c|c|}\hline
before gauge fixing & after gauge fixing \\ \hline\hline
$A_3 = A_{33} + a_3$ & $ A_3^\prime = 0 + a_3^\prime$ \\ \hline
$A_2 = A_{23} + A_{22} + a_2 $ & $ A_2^\prime = A_{23}^\prime + 0
+ a_2^\prime$ \\ \hline
$A_1  =  A_{13}  +  A_{12}  +  A_{11}  +  a_1$  &  $A_1^\prime  =
A_{13}^\prime + A_{12}^\prime + 0 + a_1^\prime$ \\ \hline
\end{tabular}

\vspace{.5cm}

{\it Table 1:  The gauge transformations to Palumbo gauge. In the
right column,  the prime denotes  successive  application  of the
transformations $U_3$, $U_2$ and $U_1$, respectively.}

\end{center} 

The explicit form of the gauge transformations $U_r$, $r= 1,2,3$,
is

\bea
U_r(x_r ) &=& P_r \exp \left[-ig  \int_{-L}^{x_r}  dy_r  (A_{rr}  + a_r
)\right] \exp\left[ ig (x_r + L) a_r^\prime \right] \nn \\[5pt]
&\equiv&  h_r[x_r  ; A_r]  \exp\left[  ig  (x_r  + L)  a_r^\prime
\right] \; .
\eea

The first,  path-ordered  exponential,  $h_r$, gauges  away ``too
much'',  namely  $A_{rr}$  {\em  and} the ZM $a_r$  and therefore
cannot be (and is not) periodic  in $x_r$.  Periodicity  and a ZM
$a_r^\prime$   are  then  restored   by  the  second  exponential
\cite{FNP81, LNT94}.  It should be noted that the ZM $a_r^\prime$
differs from the original one, $a_r$ due to (\ref{ZMWILSON}). One
has

\be
a_r^\prime = \frac{i}{2gL} \ln h[x_r = L ; A_r] \; ,
\ee

which is, however, formal in so far as the logarithm  on the {\it
r.h.s.}~is  a multivalued  function.  This raises the question of
uniqueness which will be adressed shortly when we have calculated
the FP determinant.

\ind
Before  we come  to that, let us write  down  Gauss's  law in our
decomposed notation.  The spatially constant part ($r$=0) will be
discussed  elsewhere (see also Palumbo's discussion  \cite{Pal86,
Pal90}). For $r>0$, Gauss's law reads in components

\be
G_r = d_r E_{rr} \equiv \partial_r  E_{rr} + ig [a_r , E_{rr}]  =
\FP_r [a_r] E_{rr} =  \rho_r \; .
\ee

This should  be solved  for the components  $E_{rr}$,  which  are
canonically conjugate (in the Poisson bracket sense) to the gauge
conditions $\chi_r = A_{rr}$.  To this end, one has to invert the
FP operator \cite{Pau95},

\be
\FP [a_r] = \left( \begin{array}{ccc}
                  d_1[a_1]       & 0             & 0 \\
	          ig \ad A_{12}  & d_2 [a_2]     & 0 \\
	          ig \ad A_{13}  & ig \ad A_{23} & d_3 [a_3] 
		  \end{array} 
 	     \right) \; ,
\ee	   

where the operation `ad' has been defined in (\ref{CODERIV}). The
FP operator is field dependent and in a path integral formulation
\'{a} la Faddeev and Popov would lead to ghosts. Nevertheless, it
is still much simpler than the Coulomb gauge expression,  because
it is an ordinary  rather  than a partial  differential  operator
(for each $r$). Due to the algebraic property,

\be
\partial_r a_r = 0 = [\partial_r , a_r] \; ,
\ee

the eigenvalue problem for $d_r$ furthermore separates into space
and    color.     As    a   result,    it   is   exactly    ({\it
i.e.}~non-perturbatively)  solvable  \cite{Yab89,  Pal90, LNT94}.
The eigenvalue problem (for $SU$(2)) is given by the expression

\be
\FP^{ab} u_{n_r , \alpha_r}^b  = (\partial_r \delta^{ab} + g a_r^c
\epsilon^{ab}_c   )  u_{n_r  ,  \alpha_r}^b   =  \lambda_{n_r   ,
\alpha_r}^a \; .
\ee

The eigenfunctions are found to be 

\be
u_{n_r , \alpha_r} = \exp (i\pi n_r x_r /L) w_{\alpha_r} \; ,
\ee

and separate  into a plane  wave times a color  matrix  $w$.  The
eigenvalues are

\be
\lambda_{n_r  , \alpha_r} = i \frac{\pi n_r}{L} + g \alpha_r \; ,
\quad n_r = \pm 1, \pm 2, \ldots  \neq 0 \; , \quad \alpha_r  \in
\Bigl\{0 , \pm i |a_r| \Bigr\} \; , 
\ee

where $|a_r| = (a_r^a  a_r^a )^{1/2}$  is the length of the color
vector with components $a_r$.

In  order  to  dicuss  the  uniqueness  of the  gauge  fixing, we
calculate the FP determinant

\bea
J_r &\equiv& \det \FP (a_r) / \det \FP (0)  
= \prod_{n_r , \alpha_r} \lambda_{n_r  , \alpha_r} \bigg/ \prod_{n_r
, \alpha_r} \lambda_{n_r , \alpha_r = 0} = \nonumber \\
&=& (g |a_r| L )^{-2} \sin^2 (g |a_r| L ) \; .
\eea

It is thus  essentially  given  by the  Haar  measure  of $SU$(2)
\cite{Yab89, Pal90, LNT94} and vanishes for configurations  $a_r$
of length

\be
|a_r| = \pi n_r / gL \; ,
\ee

which are gauge equivalent to $a_r = 0$. Thus one has an infinity
of Gribov copies labelled by the integers $n_r$.  The fundamental
modular  region  is therefore  given  by a ``ball''  $B$ in color
space,

\be
B: \quad 0 \le |a_r| < \pi/ gL \; .
\ee

In the region  $B$, Gauss's  law can be solved uniquely  with the
help of the inverse of $\FP$, a Green function

\be
\FP_r^{-1}  (x_r  , y_r  ) = \sum_{n_r  \neq   0 \atop \alpha_r}
\frac{u_{n_r  ,  \alpha_r}^\dagger  (x_r  )  u_{n_r  ,  \alpha_r}
(y_r)}{i\pi n_r / L + g \alpha_r} \; .
\ee

Our Hilbert space of states consists of the functionals

\be
\CH \ni \Psi [A_{13} , A_{23} , A_{12} , a_3 , a_2 , a_1 ] \equiv
\Psi_{\chi} [A^*] \; ,
\ee

where $A_{13}$, $A_{23}$ can be viewed as transverse gluons (with
the  ZMs  removed)   and  all  the  other   fields   are  ZMs  or
lower-dimensional fields \cite{LNT94}. The state functionals have
to vanish at the boundary of the fundamental region $B$,

\be
\Psi \Bigl[|a_r| = 0 \Bigr] = \Psi\Bigl[|a_r| = \pi/gL \Bigr] = 0
\; . 
\ee

The  quantum  Hamiltonian  is  implicitly  defined  on the  gauge
invariant states via

\be
H \Psi_\chi [A^*] = \bra A | H | \Psi \ket_{A_{rr} = 0} \; 
\ee

and explicitly found to be (integrating over the torus $T$),

{\arraycolsep1pt
\bea
H = \frac{1}{2}  \int\limits_T  d^3 x \,\Biggl\{  &-& \sum_{r=  0
\atop  r \neq  i}^3  J_r^{-1}  \frac{\delta}{\delta  A_{ir}}  J_r
\frac{\delta}{\delta A_{ir}} + \left.  \sum_{r=0}^3 B_{ir} B_{ir}
\right\vert_{A_{rr} = 0} + \nonumber \\
&+& \sum_{r=0}^3  J_r^{-1} (\FP_r^{-1}  * \rho_r) J_r (\FP_r^{-1} *
\rho_r) \Biggr \} \; .
\eea
}

The  first  term  containing  the functional  derivatives  is the
kinetic  energy,  the term in the second line is the analogue  of
the Coulomb terms stemming from the solution  of Gauss's law, the
asterisk  $(*)$  denoting  convolution.   As was the case for the
Coulomb gauge, the appearance  of the Jacobians  $J_r$ guarantees
the hermiticity of the Hamiltonian in the physical Hilbert space.
This in turn ensures  the unitarity  of the $S$-matrix.   We thus
note the following rule. If the FP matrix is field dependent, the
(quantum) Hamiltonian contains non-trivial  Jacobian factors.  In
the path integral  formulation  there appear ghosts.   Either  of
these features is necessary to maintain unitarity  of the theory.
This adds another  point of view to the debate whether  continuum
axial gauges, $A_3 = 0$, are ghost-free,  or contain  ghosts upon
infrared regularisation  \cite{BLS84, Nak82}.  In our opinion, it
is  still  an unsettled  problem  whether  there  is a consistent
(unitary)  formulation  of  axial-gauge  QCD  that  is definitely
ghost-free.

\vspace{1cm}

\section{Hamiltonian Yang-Mills Theory: Front-Form}

\vspace{1cm}

In the light-cone (LC) formulation  of quantum field theory based
on Dirac's front-form of relativistic  dynamics \cite{Dir49}  one
introduces  new variables  $x^\pm  = 2^{-1/2}(x^0  \pm x^3)$  and
specifies  canonical  commutators  on  hyperplanes   $x^\p  = 0$,
tangent to the light-cone  (for recent reviews see \cite{Per94}).
This  drastically   alters  the  canonical  structure   of  field
theories.   They  typically  have  Lagrangians  linear  in the LC
velocity;  some  variables  become  redundant  via  second  class
constraints and can in principle be eliminated from the theory. A
Lagrangian  linear  in the velocity  is singular  in the sense of
Dirac  \cite{Dir64}  and might  be treated  with  his theory  for
constrained systems.  More appropriate, however, is the method of
Faddeev and Jackiw \cite{FJ88}, which is explicitly  taylored for
this case.  It is essentially  equivalent to Schwinger's  quantum
action  principle  \cite{Sch51},  which  will be shown  elsewhere
\cite{Hei95}.

\ind
For  our  formulation   of  LC  Yang-Mills  theory  it  is  again
convenient  to work a in finite volume, $-L \le x^\m , x_i \le L$,
$i = 1,2$,  and  impose  pBC  in every  spatial  direction.   The
Lagrangian is

\bea
\CL &=& \Pi^a  \dot  A_\m^a  + \Pi_i^a  \dot A_i^a  - \frac{1}{2}
(\Pi^a  \Pi^a  + B^a B^a ) + A_\p^a  (D_\m^{ab}  \Pi^b + D_i^{ab}
\Pi_i^b ) \nonumber \\
&\equiv&  \Pi^a \dot A_\m^a  + \Pi_i^a  \dot A_i^a - \CH + A_\p^a
G^a \; ,
\label{LCLAG}
\eea

where $\CH$ denotes  the Hamiltonian  (density),  $G^a$ the Gauss
operator and the dotted variables derivatives with respect to the
LC time $x^\p$.  Furthermore, we have defined the chromo-magnetic
fields

\be
B = F_{12} = \partial_1 A_2 - \partial_2 A_1 + g [A_1 , A_2 ] \;
\ee

and the momenta
   
\bea
\Pi^a &=& \dot A_\m^a - D_\m^{ab} A_\p^b \; , \\
\Pi_i^a &=& \partial_\m A_i^a - D_i^{ab} A_\m^b \; . 
\label{Pii}
\eea

The Lagrangian  (\ref{LCLAG})  appears to be perfectly canonical.
However,  as  is typical  for  the  LC formulation,  the  momenta
$\Pi_i^a$  are  {\em  dependent}  variables  as (\ref{Pii})  is a
(second class) constraint, which expresses the $\Pi_i^a$ in terms
of other degrees of freedom. The number of degrees of freedom per
space-time  point  is  thus  only  12  versus  18 in the  instant
formulation.  Upon inserting (\ref{Pii}) into (\ref{LCLAG}),  the
Lagrangian becomes highly non-canonical.  Nonetheless, the method
of Faddeev and Jackiw \cite{FJ88}  can still be applied and gives
the following elementary brackets (in condensed notation)

\bea
\pb{A_i}{A_j} &=& \sfrac{1}{2} \delta_{ij} D_\m^{-1} \; , \\
\pb{A_i}{\Pi} &=& - D_i D_\m^{-1} \; , \\
\pb{A_\m}{\Pi} &=& 1 \; , \\
\pb{\Pi}{\Pi} &=& D_i D_\m^{-1} D_i \; .
\eea

These     brackets     look    highly     non-canonical     ({\it
i.e.}~non-cartesian,  as  they  differ  from  zero  or  one)  and
furthermore  are field dependent.   The inverse  of the covariant
derivative  $D_\m [A_\m]$  is difficult  to obtain  before  gauge
fixing,  even  classically.    Finally,   the  program  of  first
quantising (preferrably in cartesian coordinates) and then fixing
the gauge, as was done in the instant  formulation,  seems rather
hopeless  to be pursued.   One does not even know how to define a
(functional)  Schr\"odinger  picture.   The only possible  way to
proceed  seems to gauge fix before  quantisation,  as was already
done by Franke et al.~\cite{FNP81}.

\ind
It is illuminating  to note that all the problems above disappear
in the  pure  continuum  LC gauge,  $A_\m  = 0$, which  was  used
successfully  in many perturbative  applications (see \cite{BP91}
and references therein). In this case, the redundant momenta are

\be
\Pi_i^a = \partial_\m A_i^a \; ,
\ee

and  the  elementary  brackets  are reduced  to the canonical  LC
bracket \cite{HW94},

\be
\pb{A_i}{A_j} = \sfrac{1}{2} \delta_{ij} \partial_\m^{-1} \; .
\ee

As in the case of the axial gauge, however, the LC gauge does not
exist  on  a  torus.   This  was  already   noted  by  Franke  et
al.~\cite{FNP81}  who suggested a modification which we will call
the  FNP gauge.   It is convenient  to change  to the Cartan-Weyl
basis for $su$(2) matrices via

\be
F^a T^a  = F^\p  T^\m  + F^\m T^\p  + F^3 T^3  \; , \quad
T^a = \tau^a / 2 \; . 
\ee

The FNP gauge is then defined as

\bea
A_\m^\pm &=& 0 \; \, \nonumber \\
A_\m^3 &=& v \; , \quad \partial_\m v = 0 \; ,
\eea

where  the upper indices  refer to color,  and the lower ones are
spatial.   Again, as in the Palumbo  gauge, a ZM (with respect to
$x^\m$) is retained which in addition is diagonal in color space.
(An analogous  gauge  fixing  in the equal-time  case was used in
\cite{Yab89,  LNT94}.) We do not use the Palumbo gauge within the
front-form dynamics as this amounts to treating the indices $\m ,
i = 1, 2$ in a symmetric  fashion \cite{Pau95,  Hei95b}.   As the
fields $A_\m$, $A_i$, after elimination of the $\Pi_i$, enter the
Lagrangian  in a manifestly  asymmetric  way,  this  seems  to be
somewhat inappropriate.

\ind
Setting $G^a = 0$ for the moment, the Lagrangian becomes

\be
\CL = \pi \dot  v + A_i^\p  (\partial_\m  - igv)  \dot  A_i^\m  +
A_i^\m (\partial_\p  + igv) \dot A_i^\p + A_i^3 \partial_\m  \dot
A_i^3 - H[ \Pi^\pm , \pi, A_i] \; , \ee

from  which  one can read  off \cite{FJ88,  HW94}  the elementary
brackets

\bea
\pb{v}{\pi} &=& 1 \; ,  \\
\pb{A_i^3}{A_j^3} &=& \sfrac{1}{2} \delta_{ij} \partial_\m^{-1} \; , \\
\pb{A_i^\pm}{A_j^\mp}     &=&    -    \sfrac{1}{2}    \delta_{ij}
(\partial_\m \pm igv)^{-1} \; . 
\eea

These  lead  in the standard  manner  to the following  FP matrix
elements,

\bea
\FP_\pm (v) &=& \partial_\m \pm igv \; , \\
\FP_3 &=& \partial_\m \; .
\eea

Note that only the first two of them are field dependent  (on the
ZM $v$). Gauss's law, 

\be
G = \FP * \Pi - D_i \Pi_i = 0 \; ,
\ee

can again be solved exactly and uniquely if the FP determinant

\be
J \equiv |\det \FP| = \prod_{\alpha  = 0 , \pm 1} (\frac{\pi n}{L}
+ \alpha gv) \sim \sin^2 (gLv) \; 
\ee

is non-vanishing.   This is guaranteed in the fundamental modular
region 

\be
0 \le v < \pi/gL \; ,
\ee

which in the FNP gauge is a ``line-segment'' in color space.  The
associated  Hilbert  space in the quantum  theory can be found as
follows.  The  off-diagonal  components  $A_i^\pm$  have  a  Fock
expansion   with   $v$-dependent    coefficients,   whereas   the
coefficients  in the  expansion  of $A_i^3$  are  $v$-independent
\cite{FNP81,  KPP95}.  The canonically  conjugate  variables $v$,
$\pi  =  -i  \delta  /  \delta  v$  act  on  a Hilbert  space  of
functionals $\Psi[v]$ depending on the ZM $v$.  This implies that
the total Hilbert  space is generated  by a basis  consisting  of
products  of Fock states (associated  with the transverse  gluons
$A_i$)  with  the  functionals  $\Psi[v]$.  The  field  dependent
coefficients in the expansion of $A_i^\pm$ thus are no problem as
they act multiplicatively  ({\it i.e.}~like  pure numbers) in the
$v$-representation.   In summa, this supplements  the successfull
Fock space picture  of the continuum  LC gauge with ZM degrees of
freedom  which  are  generally  believed  to  be  connected  with
non-trivial topology \cite{FNP81, HKW91, KPP94}.

\ind
For the quantum Hamiltonian one finds

{\arraycolsep1pt
\bea
H = \frac{1}{2}  \int\limits_T  d^3 x \, \Biggl\{  &-& J^{-1} (v)
\frac{\delta}{\delta  v} J(v) \frac{\delta}{\delta v} + B^a B^a +
\nonumber \\
&+&  2 \, \FP_\m^{-1}  * (D_i  \Pi_i)^\p  \,  \FP_\p^{-1}  * (D_i
\Pi_i)^\m  + \partial_\m^{-1}  (D_i \Pi_i)^3  \, \partial_\m^{-1}
(D_i \Pi_i)^3 \Biggr\} \; ,
\label{LCHAM}
\eea
}

where for the redundant momenta $\Pi_i$ one has to substitute the
constraint  (\ref{Pii})  in the FNP gauge.  This Hamiltonian  has
been  obtained  by  Franke  et  al.~\cite{FNP81},  who,  however,
overlooked  the necessity  for the Jacobian  $J$ to appear in the
kinetic term. Unfortunately, the Hamiltonian (\ref{LCHAM}) is not
the end of story.  It still contains  redundant  variables  which
have nothing to do with a (residual) gauge freedom but are second
class constrained  and analogous to a constrained  ZM extensively
studied  in LC $\phi^4$  theory \cite{HKS92}.   The
constraint is

\be
\frac{\delta H}{\delta a_i^3} = 0 \; ,
\ee

where $a_i^3$  is the ZM of $A_i^3$ with respect  to $x^\m$.   It
cannot be solved in closed form \cite{FNP81}.  At the moment, the
implications of this constraint are unclear.

\newpage

\leftline{\Large \bf Acknowledgements}

\vspace{.5cm}

The author is indebted  to E.~Werner  for his continuing  support
and interest. He thanks A.~Bassetto, M.~Faber, J.~Fuchs, A.~Neveu
and D.~Sch\"utte  for valuable  comments and suggestions  and the
organisers  of the 35th Schladming  Winter  School  for all their
efforts.  Discussions with colleagues G.~Fischer and T.~Pause are
gratefully acknowledged.


\begin{thebibliography}{100}

\baselineskip 14pt

\bibitem{YM54}
C.N.~Yang, R.~Mills, Phys.~Rev.~{\bf 96}, 191 (1954)

\bibitem{Ber49}
P.G.~Bergmann, Phys.~Rev.~{\bf 75}, 680 (1949)\\
P.G.~Bergmann,  J.H.M.~Brunings,  Rev.~Mod.~Phys.~{\bf  21},  480
(1949)\\
P.G.~Bergmann, R.~Schiller, Phys.~Rev.~{\bf 89}, 4 (1953)

\bibitem{Dir50}
P.A.M.~Dirac, Canad.~J.~Math.~{\bf 2}, 129 (1950)

\bibitem{Dir64}
P.A.M.~Dirac,  {\sl Lectures on Quantum Mechanics}, Benjamin, New
York, 1964

\bibitem{Man62}
S.~Mandelstam, Ann.~Phys.~(N.Y.) {\bf 19}, 1 (1962)

\bibitem{Man68}
S.~Mandelstam, Phys.~Rev.~{\bf 175}, 1580 (1968)

%
%
%
%
\bibitem{Wil74}
K.G.~Wilson, Phys.~Rev.~{\bf D10}, 2445 (1974)

\bibitem{Man79}
S.~Mandelstam, Phys.~Rev.~{\bf D19}, 2391 (1979)

\bibitem{Lol93}
R.~Loll, Nucl.~Phys.~{\bf B400}, 126 (1993) \\
%
N.J.~Watson, Phys.~Lett.~{\bf B323}, 385 (1994) 

\bibitem{FSS94}
J.~Fuchs,   M.G.~Schmidt,   C.~Schweigert,   Nucl.~Phys.~{\bf   B
426},107 (1994) \\
%
J.~Fuchs,  {\sl  The  Singularity  Structure  of  the  Yang-Mills
Configuration Space}, preprint NIKHEF 95-026, hep-th/9506005

\bibitem{vBa92}
P.~van Baal, Nucl.~Phys.~{\bf B369}, 259 (1992)

\bibitem{Gri78}
V.N.~Gribov, Nucl.~Phys.~{\bf B139}, 1 (1978)

\bibitem{Sin78}
I.M.~Singer, Comm.~Math.~Phys.~{\bf 60}, 7 (1978) 

\bibitem{GB50}
S.N.~Gupta,  Proc.~Phys.~Soc.~(London)  {\bf A 63}, 681 (1950) \\
K.~Bleuler, Helv.~Phys.~Acta {\bf 23}, 567 (1950)

\bibitem{Sch62a}
J.~Schwinger,  Phys.~Rev.~{\bf  125}, 1043 (1962), {\bf 127}, 324
(1962)

\bibitem{Fey76}
R.P.~Feynman, in:  {\sl Weak and Electromagnetic  Interactions at
High Energy}, R.~Balian, C.H.~Llewellyn Smith, eds., Les Houches,
Session XXIX, 1976, North Holland, Amsterdam, 1977

\bibitem{FP67}
L.D.~Faddeev, V.N.~Popov, Phys.~Lett.~{\bf B25}, 29 (1967)

\bibitem{Fey63}
R.P.~Feynman, Acta Phys.~Polonica {\bf 26}, 697 (1963)

\bibitem{DWi67}
B.S.~De Witt, Phys.~Rev.~{\bf 162}, 1195, 1239 (1967)

\bibitem{BBR91}
D.~Birmingham, M.~Blau, M.~Rakowski, G.~Thompson, Phys.~Rep.~{\bf
209}, 129 (1991)

\bibitem{Mig75}
A.A.~Migdal,   Zh.~Eksp.~Theor.~Fiz.~{\bf    69},   810   (1975),
translated in: Sov.~Phys.~JETP {\bf 42}, 413 (1975) \\
E.~Witten, Comm.~Math.~Phys.~{\bf 141}, 153 (1991) \\
S.V.~Shabanov, Phys.~Lett.~{\bf B318}, 323 (1993) \\
L.~Chandar, E.~Ercolessi, Nucl.~Phys.~{\bf B426}, 94 (1994)

\bibitem{CL80}
N.~Christ, T.D.~Lee, Phys.~Rev.~{\bf D22}, 939 (1980)\\
T.D.~Lee,  {\sl  Particle  Physics  and  Introduction   to  Field
Theory}, Harwood, Chur, 1981, Chapter 18 

\bibitem{Pro82}
L.V.~Prokhorov, Sov.~J.~Nucl.~Phys.~{\bf 35}, 129 (1982)\\
%
K.~Kucha\v{r}, Phys.~Rev.~{\bf D34}, 3031, 3044 (1986)\\
%
L.V.~Prokhorov, S.V.~Shabanov, Phys.~Lett.~{\bf B216}, 341 (1989)
\\
%
H.~Yabuki, Ann.~Phys.~(N.Y.) {\bf 209}, 231 (1991)
%

\bibitem{CT87a}
H.~Cheng, E.-C.~Tsai, Chin.~J.~Phys.~{\bf 25}, 95 (1987)

\bibitem{Dir49}
P.A.M.~Dirac, Rev.~Mod.~Phys.~{\bf 21}, 392 (1949)

\bibitem{tHo79}
G.~`t   Hooft,   Nucl.~Phys.~{\bf   B153},   141   (1979);   Acta
Phys.~Austriaca, Suppl.~{\bf 22}, 531 (1980)

\bibitem{Lue83}
M.~L\"uscher, Nucl.~Phys.~{\bf B219}, 233 (1983)

\bibitem{FS80}
L.D.~Faddeev,  A.A.~Slavnov,  {\sl Gauge Fields:  Introduction to
Quantum Theory}, Benjamin/Cummings, Reading, 1980

\bibitem{FJ88}
L.D.~Faddeev, R.~Jackiw, Phys.~Rev.~Lett.~{\bf 60}, 1692 (1988)

\bibitem{Jac80}
R.~Jackiw, Rev.~Mod.~Phys.~{\bf 52}, 661 (1980)

\bibitem{Wei83}
W.I.~Weisberger,   in:   {\sl  Asymptotic   Realms  of  Physics},
A.H.~Guth et al., eds., MIT Press, Cambridge, Mass., 1983

\bibitem{LM94}
M.~Lavelle, D.~McMullan, Phys.~Lett.~{\bf B329}, 69 (1994) 

\bibitem{LM93}
M.~Lavelle, D.~McMullan, Phys.~Rev.~Lett.~{\bf 71}, 3758 (1993)

\bibitem{FS82}
M.A.~Semenov-Tyan-Shanskii,    V.A.~Franke,   Zapiski   Nauchnykh
Seminarov Lenin\-gradskogo Otdeleniya Matematicheskogo  Instituta
im.~V.A.~Steklov  AN SSSR {\bf 120}, 159 (1982);  translated  in:
J.~Sov.~Math.~{\bf 34}, 1999 (1986)

\bibitem{vBa94}
P.~van Baal, B.~van den Heuvel, Nucl.~Phys.~{\bf B417}, 215 (1994)

\bibitem{Khr69}
I.B.~Khriplovich,  Yad.~Fiz.~{\bf 10}, 409 (1969), translated in:
Sov.~J.~Nucl.~Phys.~{\bf 10}, 235 (1970) \\ 
%
S.D.~Drell,  in:  {\sl  A  Festschrift  for  Maurice  Goldhaber},
Transactions  of the New York Academy  of  Sciences,  Series  II,
Vol.~{\bf 40}, 76 (1980) \\
%
K.~Gottfried, Prog.~Part.~Nucl.~Phys.~{\bf 8}, 49 (1982)

\bibitem{KvB88}
J.~Koller, P.~van Baal, Nucl.~Phys.~{\bf B302}, 1 (1988)

\bibitem{Kum61}
W.~Kummer, Acta Phys.~Austriaca {\bf 14}, 149 (1961)

\bibitem{AF62}
R.L.~Arnowitt, S.I.~Fickler, Phys.~Rev.~{\bf 127}, 1821 (1962)

\bibitem{Lei87}
G.~Leibbrandt, Rev.~Mod.~Phys.~{\bf 59}, 1067 (1987)

\bibitem{BNS91}
A.~Bassetto, G.~Nardelli, R.~Soldati, {\sl Yang-Mills Theories in
Algebraic  Non-Covariant   Gauges:   Canonical  Quantization  and
Renormalization}, World Scientific, Singapore, 1991

\bibitem{Fre76}
J.~Frenkel,   Phys.~Rev.~{\bf   D13},  2325  (1976);   T.~Matsuki,
Phys.~Rev.~{\bf D19}, 2879 (1979

\bibitem{Sch63}
J.~Schwinger, Phys.~Rev.~{\bf 130}, 402 (1963)

\bibitem{Yao64}
Y.-P.~Yao, J.~Math.~Phys.~{\bf 5}, 1319 (1964)

\bibitem{BL78}
N.~Batakis, G.~Lazarides, Phys.~Rev.~{\bf D18}, 4710 (1978)

\bibitem{Wei81}
N.~Weiss, Phys.~Rev.~{\bf D24}, 475 (1981)\\
E.~D'Hoker, Nucl.~Phys.~{\bf B201}, 401 (1982)\\
G.~Curci, G.~Menotti, Z.~Phys.~{\bf C21}, 281 (1984)

\bibitem{FNP81}
V.A.~Franke,        Yu.~V.~Novozhilov,         E.V.~Prokhvatilov,
Lett.~Math.~Phys.~{\bf 5}, 239, 437 (1981)

\bibitem{BLS84}
A.~Bassetto, I.~Lazzizzera,  R.~Soldati, Nucl.~Phys.~{\bf  B236},
319 (1984)

\bibitem{Pal86}
F.~Palumbo, Phys.~Lett.~{\bf B173}, 81 (1986)

\bibitem{Yab89}
H.~Yabuki, Phys.~Lett.~{\bf B231}, 271 (1989)

\bibitem{Pal90}
F.~Palumbo, Phys.~Lett.~{\bf B243}, 109 (1990) 

\bibitem{LNT94}
F.~Lenz,  H.~Naus,  M.~Thies,  Ann.~Phys.~(N.Y.)  {\bf 233},  317
(1994)

\bibitem{KLS94}
E.~Langmann, M.~Salmhofer, A.~Kovner, Mod.~Phys.~Lett.~{\bf  A9},
2913 (1994)

\bibitem{Pau95}
T.~Pause, diploma thesis, Regensburg, 1995

\bibitem{Hei}
T.~Heinzl, in preparation

\bibitem{Nak82}
N.~Nakanishi,    Progr.~Theor.~Phys.~{\bf    67},   965   (1982);
Phys.~Lett.~{\bf B131}, 381 (1983) \\
%
A.~Burnel, Phys.~Rev.~{\bf D36}, 1852 (1987) \\
%
H.~Cheng, E.-C.~Tsai, Phys.~Rev.~{\bf D36}, 3196 (1987) \\
%
T.~Heinzl,  S.~Krusche,  {\sl  A  Path-Integral   Formulation  of
Yang-Mills  Theory  in  Regularised   Axial  Gauges},  Regensburg
preprint, TPR 90-16, 1990 (unpublished)

\bibitem{Sch51}
J.~Schwinger,  Phys.~Rev.~{\bf  82},  914 (1951);  {\bf 91},  713
(1953)

\bibitem{Per94}
R.J.~Perry, {\sl Hamiltonian Light-Front Field Theory and Quantum
Chromodynamics},  Invited  lectures  presented  at `Hadrons  94',
Gramado, Brasil, April 1994, hep-th/9407056 \\
M.~Burkardt,    to   appear    in   Adv.~Nucl.~Phys.~{\bf    23},
hep-ph/9505259

\bibitem{Hei95}
T.~Heinzl, in preparation

\bibitem{BP91}
S.J.~Brodsky,   H.-C.~Pauli,   {\sl  Light-Cone  Quantization  of
Quantum  Chromodynamics},  Lecture  Notes  in Physics  {\bf 396},
Proceedings,   Schladming,   Austria,  1991;  Springer,   Berlin,
Heidelberg, New York, 1991

\bibitem{HW94}
T.~Heinzl, E.~Werner, Z.~Phys.~{\bf C62}, 521 (1994)

\bibitem{Hei95b}
T.~Heinzl,  Nucl.~Phys.~B  (Proc.~Suppl.)  {\bf  39},  B,  C, 217
(1995)

\bibitem{KPP95}
A.C.~Kalloniatis, H.-C.~Pauli, S.S.~Pinsky, Phys.~Rev.~{\bf D52},
1176 (1995)

\bibitem{HKW91}
T.~Heinzl,  S.~Krusche,  E.~Werner,  Phys.~Lett.~{\bf  B256},  55
(1991)

\bibitem{KPP94}
A.C.~Kalloniatis, H.-C.~Pauli, S.S.~Pinsky, Phys.~Rev.~{\bf D50},
6633 (1994)

\bibitem{HKS92}
T.~Heinzl, S.~Krusche,  S.~Simb\"urger,  E.~Werner, Z.~Phys.~{\bf
C56}, 415 (1992) \\
%
S.S.~Pinsky, B.~van de Sande, C.M.~Bender, Phys.~Rev.~{\bf  D48},
816 (1993)\\
S.S.~Pinsky, B.~van de Sande, Phys.~Rev.~{\bf D49}, 2001 (1994)\\
S.S.~Pinsky, B.~van de Sande, J.R.~Hiller, Phys.~Rev.~{\bf  D51},
726 (1995) \\
%
T.~Heinzl,  C.~Stern, E.~Werner,  B.~Zellermann,  {\sl The Vacuum
Structure  of  Light-Front   $\phi^4_{1+1}$-Theory},   Regensburg
preprint, TPR 95-20, 1995, to appear in Z.~Phys.~{\bf C}

\end{thebibliography}
\end{document}